\documentclass[iop]{emulateapj}
\usepackage{xcolor}
\usepackage{amsmath}
\usepackage{color}
\usepackage{soul}
\begin{document}

\title{A Characteristic Transmission Spectrum dominated by H$_{2}$O applies to the majority of HST/WFC3 exoplanet observations}

\author{Aishwarya R. Iyer\altaffilmark{1,2}, Mark R. Swain\altaffilmark{1}, Robert T. Zellem\altaffilmark{1}, Michael R. Line\altaffilmark{3}, Gael Roudier\altaffilmark{1}, \\ Gra\c{c}a Rocha\altaffilmark{1}, and John H. Livingston\altaffilmark{1}}

\affil{$^{1}$ Jet Propulsion Laboratory, California Institute of Technology, 4800 Oak Grove Dr, Pasadena, CA 91109, USA}
\affil{$^{2}$ California State University, Northridge, 18111 Nordhoff St., Northridge CA 91330}
\affil{$^{3}$ Department of Astronomy and Astrophysics, University of California-Santa Cruz, 1156 High Street, Santa Cruz, CA 95064, USA}

\email{aishwarya.iyer@jpl.nasa.gov, mark.r.swain@jpl.nasa.gov}

\begin{abstract}
Currently, 19 transiting exoplanets have published transmission spectra obtained with the Hubble/WFC3 G141 near-IR grism. Using this sample, we have undertaken a uniform analysis incorporating measurement-error debiasing of the spectral modulation due to H$_{2}$O, measured in terms of the estimated atmospheric scale height, ${H_s}$. For those planets with a reported H$_{2}$O detection (10 out of 19), the spectral modulation due to H$_{2}$O ranges from  0.9 to 2.9~${H_s}$ with a mean value of 1.8~$\pm$~0.5~${H_s}$. This spectral modulation is significantly less than predicted for clear atmospheres.  For the group of planets in which H$_{2}$O has been detected, we find the individual spectra can be coherently averaged to produce a characteristic spectrum in which the shape, together with the spectral modulation of the sample, are consistent with a range of H$_{2}$O mixing ratios and cloud-top pressures, with a minimum H$_{2}$O mixing ratio of 17~$^{+12}_{-6}$~ppm corresponding to the cloud-free case. Using this lower limit, we show that clouds or aerosols must block at least half of the atmospheric column that would otherwise be sampled by transmission spectroscopy in the case of a cloud-free atmosphere. We conclude that terminator-region clouds, with sufficient opacity to be opaque in slant-viewing geometry, are common in hot Jupiters.
\end{abstract}

\keywords{Hubble Space Telescope: WFC3 --- Hot Jupiters: transmission spectrum --- H$_2$O-detection --- methods:analytical --- atmospheres --- radiative transfer --- planets and satellites: general}

\section{Introduction}

The search for H$_{2}$O in exoplanet atmospheres has been dominated by transmission measurements obtained with space-based instruments.  Although the early detections of H$_{2}$O in an exoplanet atmosphere were made with the \emph{Hubble} and \emph{Spitzer} instruments STIS, IRAC, and NICMOS \citep{barman08,tinetti07,swain08, grillmair08}, the leading instrument in this area is NASA's Hubble Space Telescope (HST) Wide~Field~Camera~3 (WFC3) using the G141 IR grism (1.1--1.7 $\mu$m), to obtain spectra of the transit event. The scope of the collective work is impressive and constitutes the largest collection, 19, of similarly-observed exoplanets presently available. These 19 transmission spectra are drawn from 16 papers by 13 authors, a majority (10 of 19) of which report a detection of H$_2$O \citep[Table~\ref{table:target_table};][]{deming13, ehrenreich14, fraine14, huitson13, knutson14a, knutson14b, kreidberg14a, kreidberg14b, kreidberg15, line13b, mandell13, mccullough14, ranjan14, sing15, wakeford13, wilkins14}. As a whole, this sample represents a heterogeneous collection of data reduction methods, spectral resolution, observational, and model interpretation approaches. Given these differences, we focus our analysis on the H$_2$O absorption feature, which occurs in the near-infrared at $\sim$ 1.2--1.6~ $\mu$m. Here we report the trends for both spectral modulation and spectral shape and discuss the possible implications of these findings.


\section{Methods}


\begin{deluxetable*}{lcccccr}
\tablecaption{HST/WFC3 IR exoplanet transmission observations used in this analysis}
\tablehead{\colhead{Object} & \colhead{T$_{eff}$} & \colhead{Planetary} & \colhead{Derived Spectral} & \colhead{Derived Spectral} & \colhead{Spectral} & \colhead{} \\
\colhead{Name} & \colhead{Calculated (K)} & \colhead{Scale Height (km)} & \colhead{Modulation (ppm)} & \colhead{Modulation (H$_{s}$)} & \colhead{Channels} & \colhead{Source}} \\
\startdata
\multicolumn{7}{c}{\textbf{H$_{2}$O-detection Reported}}\\
HAT-P-1b  &  1304$\pm 40$  &  544$\pm 58$  &  430$\pm 178$  &  2.8$\pm 1.2$  &  28 & \citet{wakeford13} \\
HAT-P-11b  &  870$\pm 16$  &  269$\pm 33$  &  127$\pm 47$  &  2.2$\pm 0.8$  &  29 & \citet{fraine14} \\
HD~189733b  &  1199$\pm 21$  &  197$\pm 14$  &  222$\pm 63$  &  2.0$\pm 0.6$  &  28 & \citet{mccullough14} \\
HD~209458b  &  1445$\pm 19$  &  558$\pm 25$  &  241$\pm 38$  &  1.5$\pm 0.2$  &  28 & \citet{deming13} \\
WASP-12b  &  2581$\pm 90$  &  951$\pm 106$  &  280$\pm 21$  &  1.5$\pm 0.1$  &  8 & \citet{kreidberg15} \\
WASP-17b  &  1546$\pm 58$  &  1000$\pm 152$  &  587$\pm 232$  &  1.5$\pm 0.6$  &  19 & \citet{mandell13} \\
WASP-19b  &  2064$\pm 46$  &  502$\pm 29$  &  306$\pm 86$  &  1.5$\pm 0.4$  &  6 & \citet{huitson13} \\
WASP-31b  &  1572$\pm 35$  &  1140$\pm 105$  &  359$\pm 270$  &  1.1$\pm 0.8$  &  25 & \citet{sing15} \\
WASP-43b  &  1374$\pm 78$  &  97$\pm 15$  &  99$\pm 66$  &  1.4$\pm 0.9$  &  22 & \citet{kreidberg14b} \\
XO-1b  &  1206$\pm 29$  &  275$\pm 31$  &  292$\pm 110$  &  2.7$\pm 1.0$  &  29 & \citet{deming13} \\
\hline \\
\multicolumn{7}{c}{\textbf{Non-H$_{2}$O-detection Reported}}\\
CoRoT-1b  &  1897$\pm 81$  &  598$\pm 95$  &  845$\pm 959$  &  4.1$\pm 4.6$  &  10 & \citet{ranjan14} \\
CoRoT-2b  &  1537$\pm 39$  &  144$\pm 10$  &  95$\pm 77$  &  1.3$\pm 1.0$  &  11 & \citet{wilkins14} \\
GJ~436b  &  649$\pm 58$  &  183$\pm 20$  &  49$\pm 44$  &  0.5$\pm 0.5$  &  28 & \citet{knutson14a} \\
GJ~1214b  &  560$\pm 29$  &  226$\pm 46$  &  16$\pm 28$  &  0.0$\pm 0.1$  &  22 & \citet{kreidberg14a} \\
GJ~3470b  &  651$\pm 55$  &  294$\pm 88$  &  3$\pm 31$  &  0.0$\pm 0.2$  &  107 & \citet{ehrenreich14} \\
HAT-P-12b  &  957$\pm 17$  &  603$\pm 47$  &  0$\pm 373$  &  0.0$\pm 1.1$  &  23 & \citet{line13b} \\
HD~97658b  &  733$\pm 23$  &  169$\pm 28$  &  22$\pm 18$  &  1.1$\pm 0.9$  &  28 & \citet{knutson14b} \\
TrES-2b  &  1497$\pm 32$  &  269$\pm 22$  &  286$\pm 162$  &  3.0$\pm 1.7$  &  10 & \citet{ranjan14} \\
TrES-4b  &  1784$\pm 40$  &  861$\pm 95$  &  524$\pm 451$  &  3.9$\pm 3.4$  &  10 & \citet{ranjan14} \\

\enddata
\label{table:target_table}
\end{deluxetable*}

\begin{figure*}
\includegraphics[width=1\textwidth]{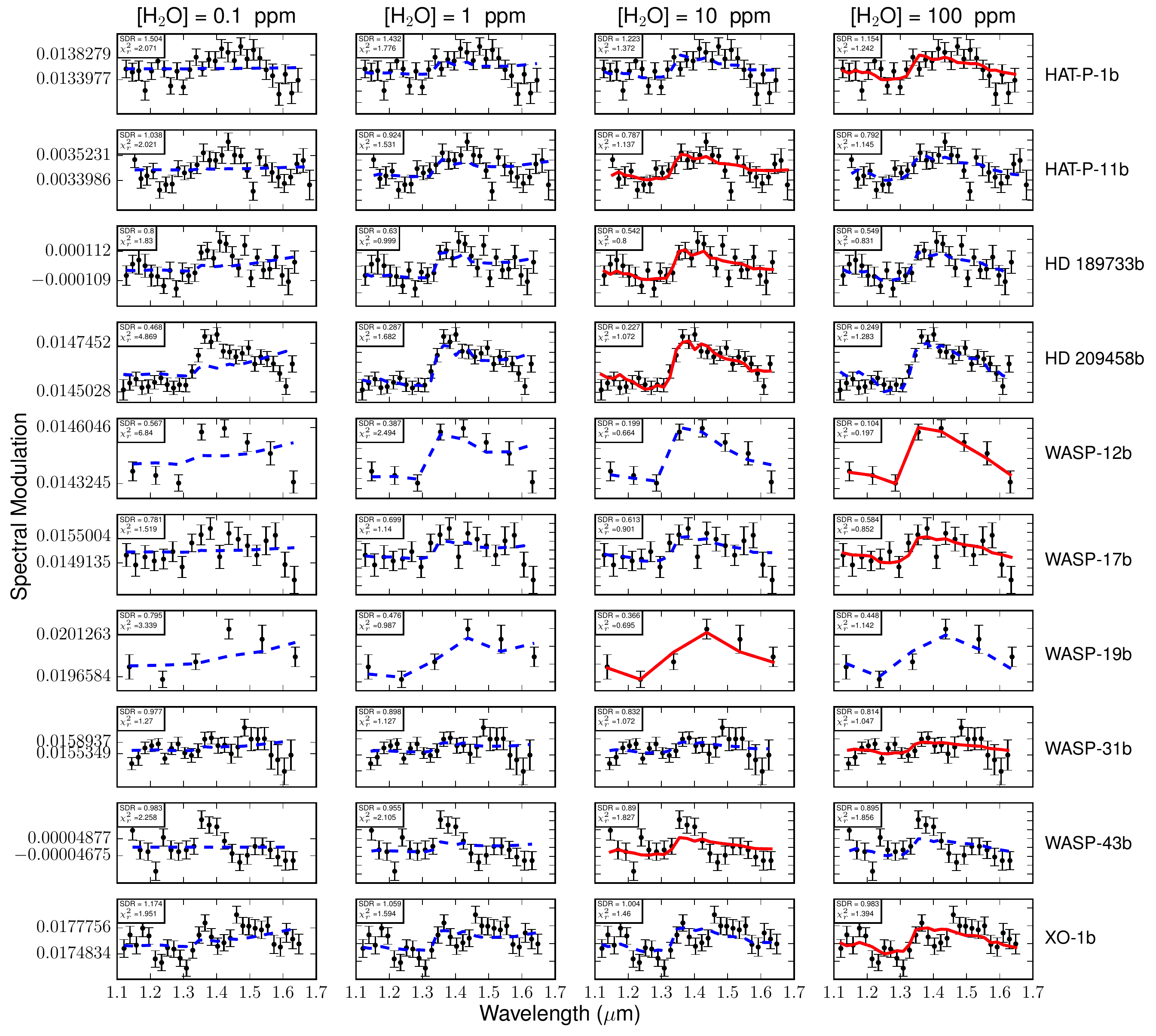} 
\caption{The 10 published HST/WFC3 IR transmission spectra of exoplanets with a H$_{2}$O detection (black points; Table~\ref{table:target_table}) are fit to a grid of cloud-free H$_{2}$O models with abundances of 0.1, 1, 10, and 100~ppm (blue dashed line) by scaling the models' offsets and amplitudes. The standard deviation of the residuals (SDR) in units of scale height and the reduced chi squared ($\chi^2_r$) are noted for each case, where the minimum $\chi^2_r$ (as well as SDR) indicates the best-fit model for each target (solid red line). This analysis allows us to debias the estimate of the spectral modulation amplitude of the H$_{2}$O feature from outlying data points.}
\label{fig:postage_stamp}
\end{figure*}

Given a heterogeneous collection of measurements presently, we adopt a template-fitting approach to determine the spectral modulation due to H$_{2}$O opacity. We define spectral modulation as the amplitude of the H$_{2}$O feature between 1.2--1.4  $\mu$m. We generate H$_{2}$O template spectra that include the opacities due to Rayleigh scattering and H$_{2}$/H$_{2}$ and H$_{2}$/He collisionally induced absorption, using the CHIMERA forward model routine for transmission spectra \citep{line13a, swain14, kreidberg14b, kreidberg15} over the spectral range of the G141 grism (1.1--1.7~$\micron$). These H$_{2}$O templates cover abundances from 0.1 to 100 ppm, the range over which the $shape$ of the H$_{2}$O spectral modulation changes. H$_{2}$O mixing ratios below 0.1~ppm are difficult to detect and abundances above 100~ppm produce spectra that have nearly the same shape when normalized. These templates are then fit to the HST/WFC3 data (Fig.~\ref{fig:postage_stamp}) with a Levenburg-Markwardt least-squares minimization routine \citep[e.g.,][]{markwardt09}, by linearly scaling the model amplitudes and vertical offsets. This exercise is carried out in order to debias the estimate for spectral modulation from single point outliers and to create a consistent method to treat data reported spectral resolutions that differ by $\sim$5.

To facilitate further analysis, all of the HST/WFC3 data and their corresponding best-fit radiative transfer models are converted to units of planetary scale height $H_{s}$: 

\begin{equation}
H_s = \frac{k_{B}T_{eq}}{\mu g}.
\end{equation} 
where $k_B$ is the Boltzmann constant, $\mu$ is the mean molecular weight of an atmosphere in solar composition ($\mu$ = 2.3 amu), $g$ is the surface gravity, and $T_{eq}$ is the calculated equilibrium temperature. We adopt the planetary parameters listed on exoplanets.org \citep{han14} for all of these variables except the equilibrium temperature. Assuming efficient heat redistribution from the dayside to the nightside and an albedo of zero, we calculate the equilibrium temperature for each planet via the equation \citep{mendez14}:
\begin{equation}
T_{eq} = \frac{T_{*}{f^{' \frac{1}{4}}}}{(a/R_{*})^{\frac{1}{2}}}\sqrt{1 + \left(\frac{8}{\pi^2}-1\right) e^{\frac{5}{2}}}
\end{equation}
where $T_*$ and $R_*$ are the stellar temperature and radius, $a$ is the semi-major axis, and $e$ is the eccentricity.

\begin{figure}
\centering
\includegraphics[width=1\columnwidth]{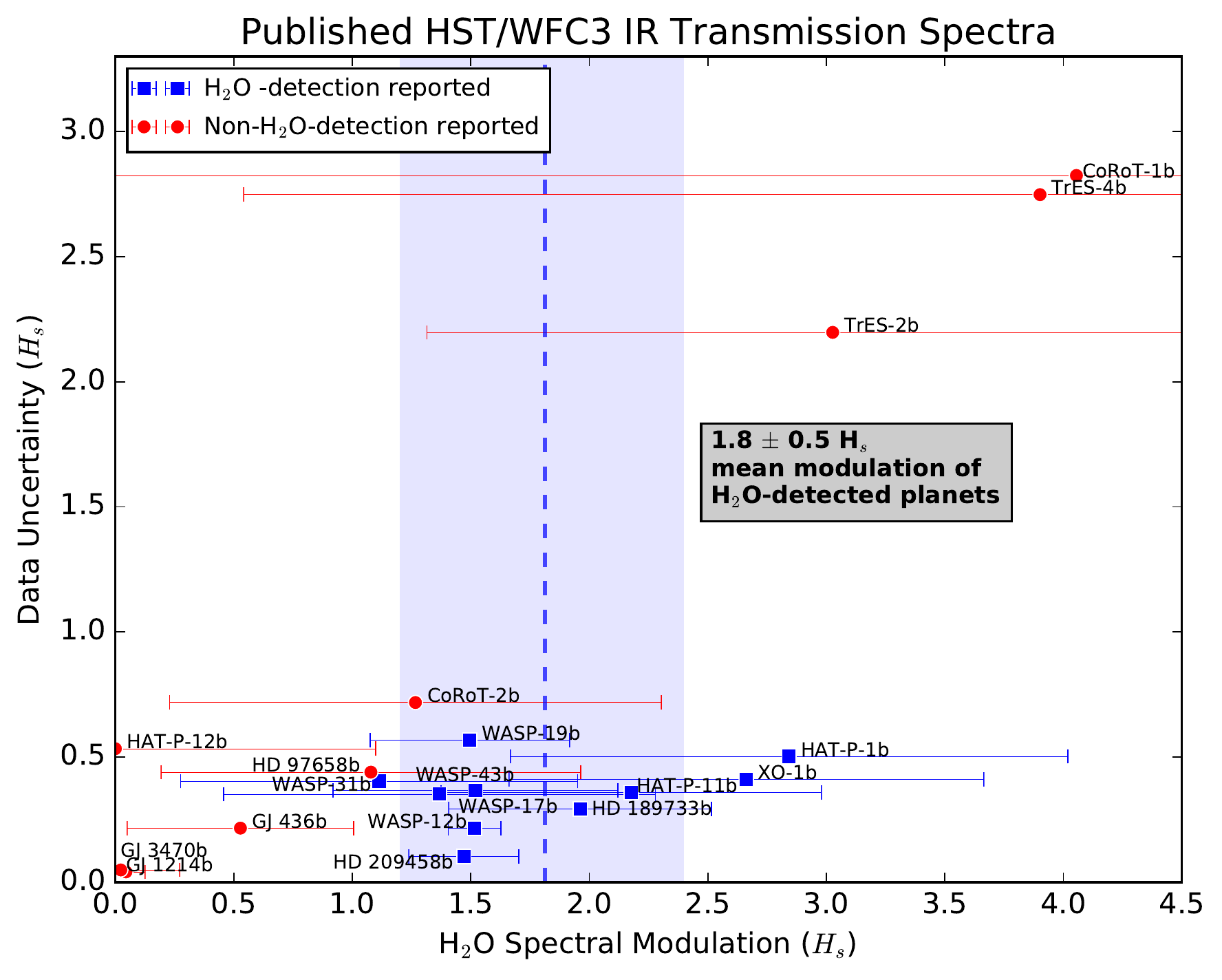} 
\caption{The H$_{2}$O spectral modulation in scale heights of the 19 published HST/WFC3+G141 transiting exoplanet transmission spectra are plotted here versus the uncertainty of the data. The horizontal error bars are standard deviation of the residuals of the cloud-free H$_{2}$O model fits (see Fig.~\ref{fig:postage_stamp}). The exoplanets with reported H$_{2}$O detections are depicted with blue squares while reported non-H$_{2}$O-detections are red circles. We find that the H$_{2}$O-detected planets (blue) have a mean spectral modulation of 1.8 $\pm$ 0.5 H$_{s}$; this value is smaller than the predicted spectral modulation for a clear atmosphere, suggesting that these planets have a cloud deck.}
\label{fig:phase_space}
\end{figure}

The modulation of each planet's H$_{2}$O feature ($\sim$1.2--1.4~$\micron$) is determined by the amplitude of the best fit model template to prevent any outliers in each dataset from skewing the spectral fit. This parameter is then plotted versus the data uncertainty, defined as the mean uncertainty in each spectra scaled by the square root of the change in resolution. The error bars on the spectral modulation are defined as the standard deviation of the residuals of the best-fit template model, (Fig.~\ref{fig:phase_space}) and all values here are in units of H$_{s}$.

The major outlying points of TrES-2b, TrES-4b and CoRoT-1b show a large uncertainty in their spectra as well as significant spectral modulation (Fig.~\ref{fig:phase_space}). However, their poor fit to the water templates indicate that caution should be used in interpreting modulation results for these planets. 



\begin{figure}
\centering
\includegraphics[width=1\columnwidth]{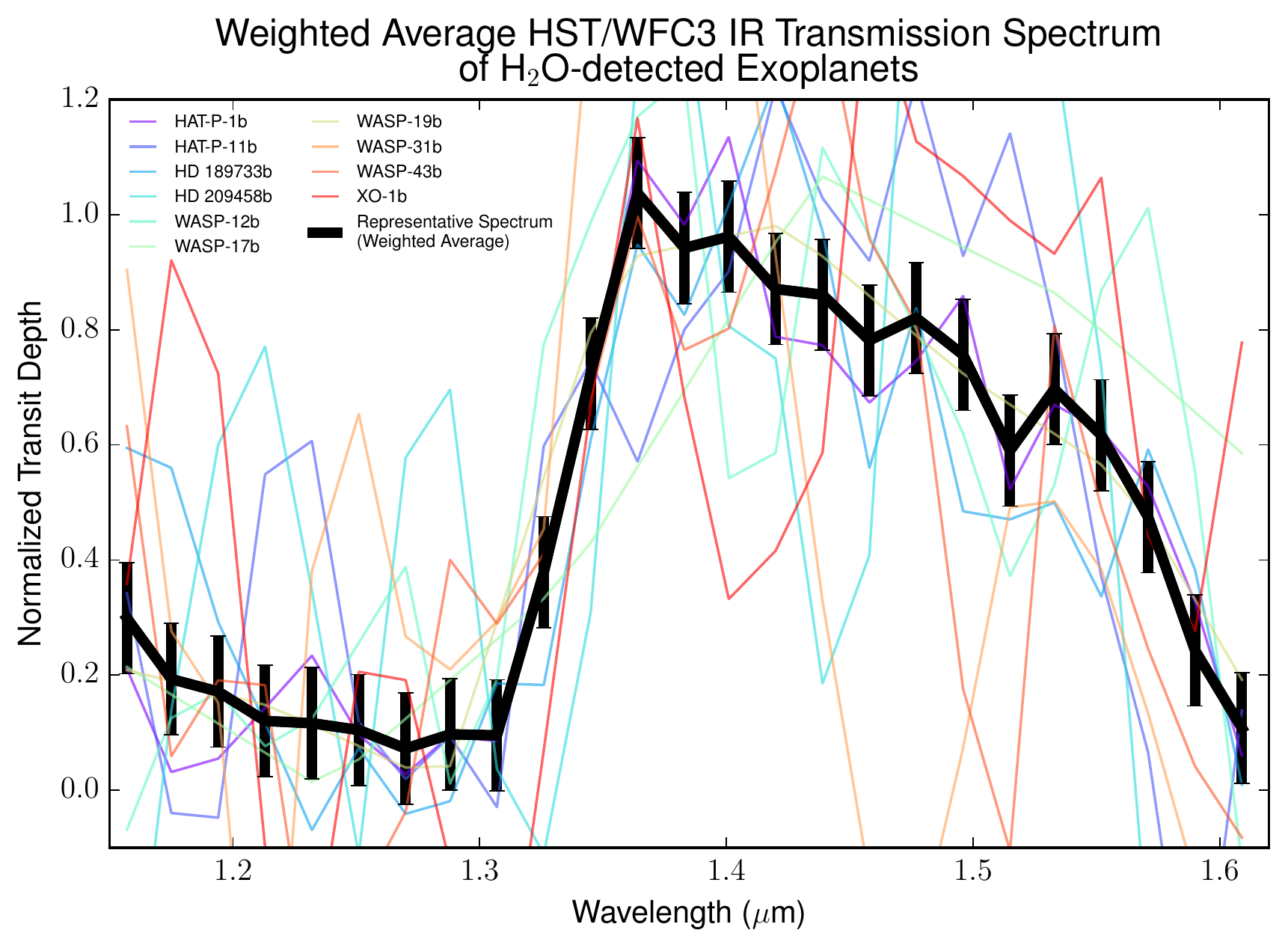}
\caption{The normalized HST/WFC3 IR transmission spectra of the 10 exoplanets with reported H$_2$O-detections (thin lines) are combined with a weighted mean to create a representative spectrum of H$_{2}$O-bearing exoplanets (thick black line).}
\label{fig:representative_spectrum}
\end{figure}

We then search for a ``representative'' spectrum shared among the exoplanets published with H$_{2}$O detections.  A similar approach to analyze \emph{Spitzer} data is used by \cite{schwartz15}. We construct a cumulative H$_{2}$O-detection transmission spectrum by normalizing the 10 published H$_{2}$O-detected spectra between 0 and 1, linearly interpolating them to a common wavelength grid, and then combining them with a weighted average. Unbiased uncertainties of this weighted average spectrum are calculated using the standard expression for the error in the weighted mean. The resultant spectrum (Fig.~\ref{fig:representative_spectrum}) has a characteristic shape that is representative of this group of HST/WFC3 planets, with H$_{2}$O as the dominating feature.


%

\section{Results and Analysis}\label{resultsandanalysis}
\begin{figure}
\centering
\includegraphics[width=1\columnwidth]{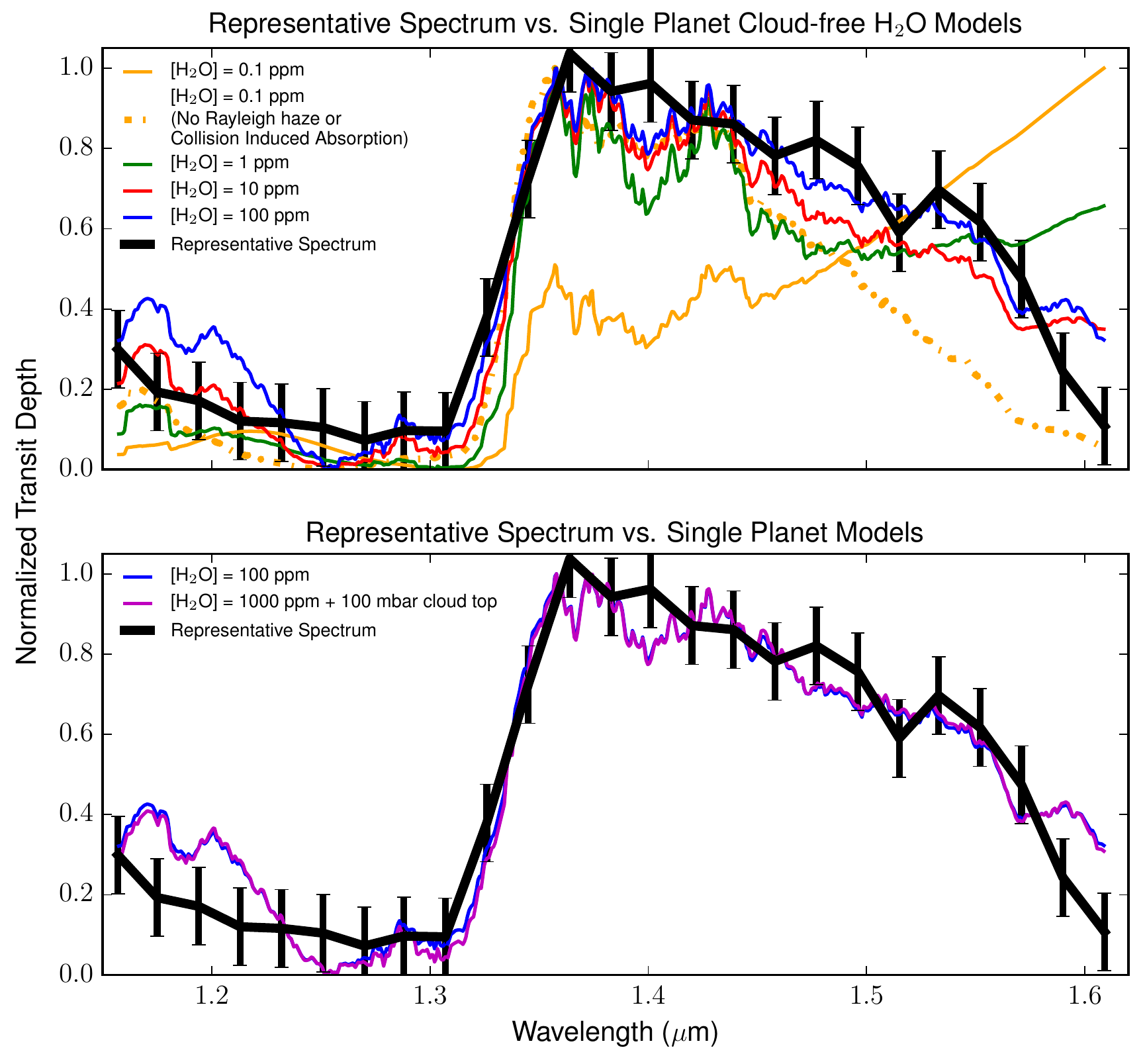}
\caption{Single planet models with Rayleigh haze slope of 1 \citep[computed using Equation 1 of][]{lecavelierdesetangs08} with the CHIMERA routine \citep{line13a, swain14, kreidberg14b, kreidberg15} and assuming H$_{2}$/H$_{2}$ and H$_{2}$/He collisionally induced absorption in addition to the following: \emph{Top:} Cloud-free forward models of varying H$_{2}$O abundances (thin multi-color lines) compared to the representative spectrum (thick black line, see Fig.~\ref{fig:representative_spectrum}). The increase in transit depth at longer wavelengths for the [H$_{2}$O] = 0.1~ppm model is due to H$_{2}$/H$_{2}$ and H$_{2}$/He collisionally induced absorption. We illustrate this effect with an [H$_{2}$O] = 0.1~ppm model without these opacity sources (yellow dashed line). Higher H$_{2}$O abundance models have enough H$_{2}$O to mask these features, yielding a similar spectral shape when normalized. \emph{Bottom:} Two forward models, one cloud-free with [H$_{2}$O]~=~100~ppm (thin blue line) and the other with [H$_{2}$O]~=~1000~ppm and a cloud top at 100~mbar (thin purple line), are compared to the representative spectrum (thick black line). These two models are nearly identical and as such fit the representative spectrum similarly well, illustrating the degeneracy of the cloud-top pressure and the water abundance.}
\label{fig:shape_analysis}
\end{figure}

To test the validity of this H$_2$O-detected ``representative" spectrum, and to understand the emerging patterns pertaining to this group, we compare it with four single-planet clear (cloud and haze-free) atmosphere models with H$_2$O abundances of 0.1, 1, 10, and 100~ppm. We include Rayleigh scattering and H$_{2}$/H$_{2}$ and H$_{2}$/He collisionally induced absorption as additional opacity sources in these models, as the planets in our sample are predominantly hot Jupiters with hydrogen and helium atmospheres. These models are also normalized between 0 and 1 to facilitate the comparison of their shape to that of the representative spectrum (Fig.~\ref{fig:shape_analysis},~\emph{Top}). We find that the amplitude of the representative spectrum is consistent with H$_2$O abundances of 10 to 100~ppm. 



Additionally, we also explore the effect of clouds on the H$_{2}$O spectral modulation amplitude to explain the shape of the representative spectrum. We compare the representative spectrum to a forward model with [H$_{2}$O]~=~1000~ppm and a cloud top at 100~mbar, alongside a cloud-free model with [H$_{2}$O]~=~100~ppm (Fig.~\ref{fig:shape_analysis}, \emph{Bottom}). The choice of H$_{2}$O mixing ratio and cloud-top pressure were selected to match best to the representative spectrum. Both models show a good qualitative fit relative to the representative spectrum indicating a degeneracy between the cloud-free and cloudy solutions.

\begin{figure*}
\centering
\includegraphics[width=0.495\textwidth]{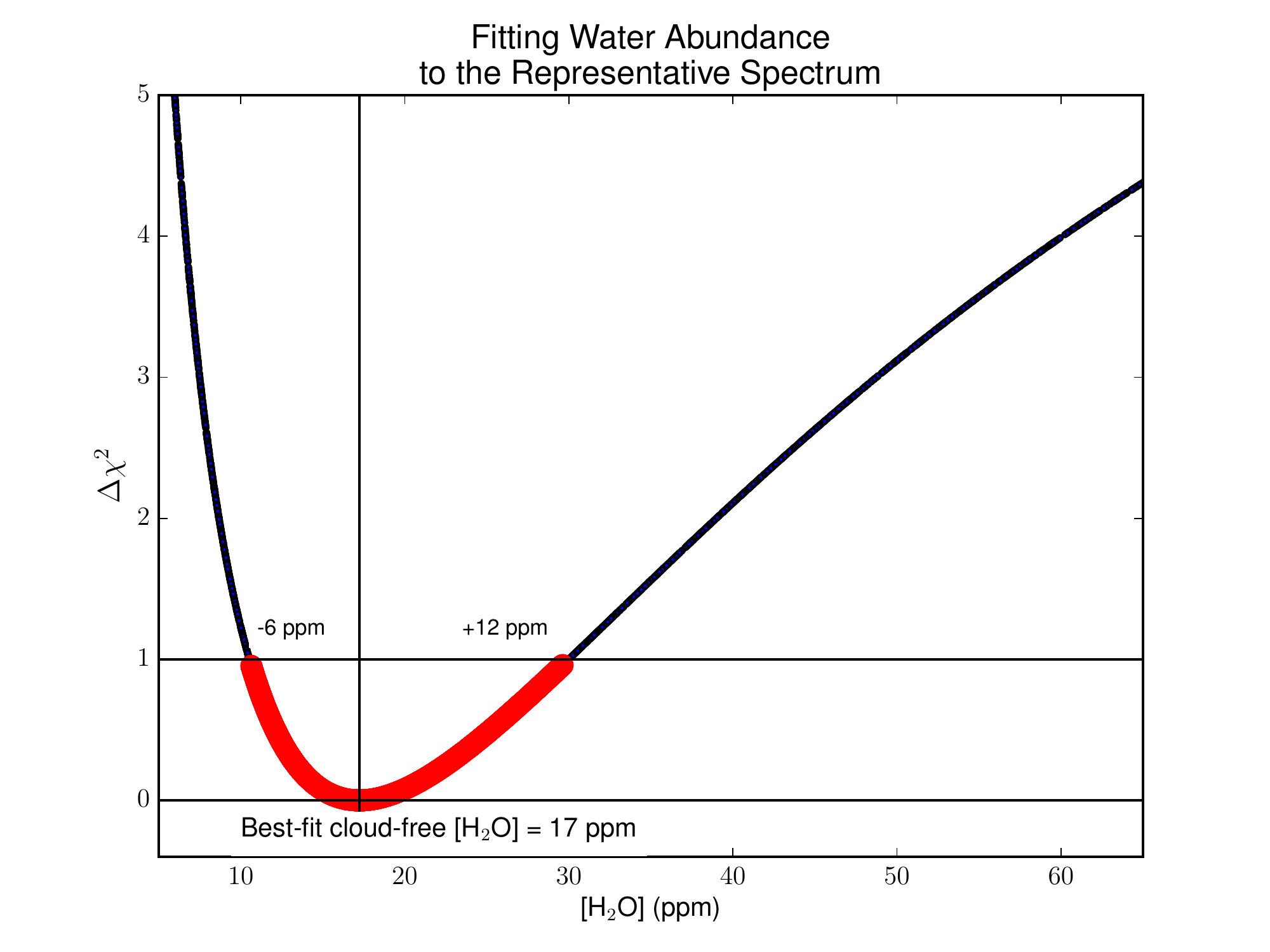} \includegraphics[width=0.495\textwidth]{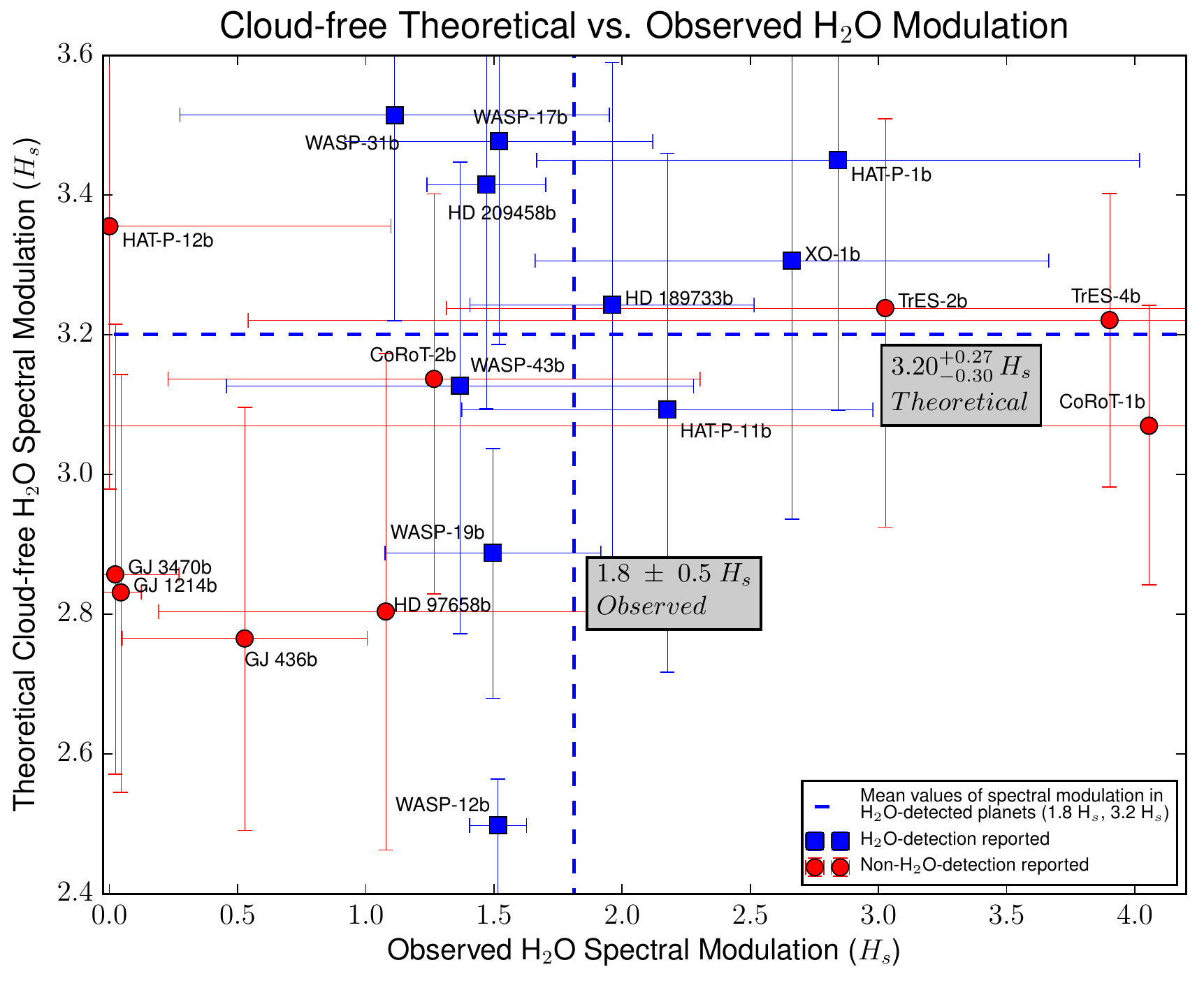}
\caption{\emph{Left:} 10,000 synthetic cloud-free spectra are generated with H$_{2}$O abundances between 1 and 1000~ppm. Each spectrum is normalized between 0 and 1 and compared to the representative spectrum (Fig.~\ref{fig:representative_spectrum}) with a $\Delta \chi^{2}$ calculation. The best fit $\Delta \chi^{2}$ between the synthetic data and the representative spectrum has a [H$_{2}$O] = 17 $^{+12}_{-6}$~ppm. \emph{Right:} Clouds typically  prevent at least half the potentially measurable atmospheric annulus from being sampled in a transit measurement. We illustrate this by computing the spectral modulation for a cloud free atmosphere using an [H$_{2}$O] abundance of 17 $^{+12}_{-6}$~ppm. Cloud-free forward models are generated for each HST/WFC3 planet and the observed H$_{2}$O spectral modulation is plotted versus the theoretical pure-H$_{2}$O modulation derived from the forward models. As a whole, the H$_{2}$O-detection reported planets (blue squares) have a mean observed modulation of 1.8$\pm 0.5$~H$_{s}$ (Fig.~\ref{fig:phase_space}) compared to a theoretical modulation of 3.20$^{+0.27}_{-0.30}$~H$_{s}$. The ratio of these values represents a maximum value ($\sim$2 H$_{s}$)  of potentially view-able atmospheric column sampled by the transmission spectrum.}
\label{fig:chi2_plots}
\end{figure*}

To explore the range of values for the water mixing ratio that are consistent with the data, we perform a $\Delta \chi^{2}$ search of the parameter space by computing forward models and comparing their shape with the representative spectrum (Fig.~\ref{fig:chi2_plots}, \emph{left}). To generate these forward models, we use parameters of a `representative planet' by computing the average T$_{eq}$, R$_{p}$, R$_{s}$ and log(g) of all the water hosting planets in our sample. Cloud-free models with cloud-top pressure of $\ge$1~bar, where they do not interact with the transmission spectrum can be consistent with the data. Larger values for the H$_{2}$O mixing ratio can also be consistent with the data, but are degenerate with cloud-top pressure \citep{benneke15, kreidberg15}. However, we can estimate the degree to which clouds block portions of the atmosphere that would otherwise be sampled in a transmission spectrum.  Using the representative spectrum's best-fit cloud-free water abundance of 17 $^{+12}_{-6}$~ppm (Fig.~\ref{fig:chi2_plots}, \emph{left}), we compute the spectral modulation for all the planets in our sample in the following way. We run 1000 Monte-Carlo iterations per planet to generate cloud free forward models with [H$_{2}$O] abundance sampled from the asymmetrical distribution (Fig.~\ref{fig:chi2_plots}, \emph{left}). Planet parameters of T$_{eq}$, R$_{p}$, R$_{s}$ and log(g) unique to each planet are used in the CHIMERA radiative transfer code \citep{line13a, swain14, kreidberg14b, kreidberg15}. We then calculate the theoretical cloud-free H$_{2}$O spectral modulation derived from the MC for each planet, which is plotted against the measured spectral modulation (Fig.~\ref{fig:chi2_plots}, \emph{right}). By averaging the results for the water-detected planets, we find that clouds likely remove at least half, and possibly more, of the observable atmospheric modulation from participating in a transit measurement.

\section{Discussions}
The H$_{2}$O spectral modulation of the HST/WFC3 H$_{2}$O-detected exoplanets spans 0.9 to 2.9~$H_{s}$ with a mean modulation of 1.8~$\pm$~0.5~$H_{s}$. This mean value differs significantly from expectations for the cloud-free H$_{2}$O spectral modulation as compared with idealized models which predict $\sim$7 scale heights of spectral modulation \citep{seager00, brown01}. This reduced H$_{2}$O modulation implies the presence of some additional opacity source, such as clouds or aerosol haze, to reduce the true spectral modulation due to H$_{2}$O.  Aerosol haze has been reported in the atmosphere of the H$_{2}$O-hosting hot Jupiter HD~189733b \citep{lecavelierdesetangs08, pont08, sing09} and clouds have been discussed by \cite{brown01} and \cite{morley12}. Quite literally, we are likely seeing the effects of H$_{2}$O above the haze/clouds which are opaque for transit viewing geometry, but may not be so for vertical paths observed during eclipse. This hypothesis agrees with recent findings for some of the planets in our sample by \citet{benneke15}, \citet{kreidberg15} and \citet{sing16}.

The presence of a cloud deck as a common feature of the hot Jupiter
H$_{2}$O-hosting planets has implications for recent work which
found a measurable difference ($\sim$100$\times$) in the H$_{2}$O
mixing ratio for the dayside and terminator regions of HD 189733b
\citep{line14, madhusudhan14}.  The spectral modulation due to H$_{2}$O in the
terminator region is 2.0 $\pm$ 0.5~$H_{s}$. In a scenario where terminator-region clouds are ubiquitous, the dayside measurement can be
understood as probing a region in which dayside perpendicular viewing
potentially samples 1.3~$H_{s}$ of additional atmospheric path in deeper, higher-pressure regions of the atmosphere. This geometry allows for more H$_{2}$O opacity to be present in the emission spectrum and a correspondingly
large value is determined for the H$_{2}$O mixing ratio.



\section{Conclusions}
Although we now possess the observational evidence to support the
conclusion that H$_{2}$O is a common constituent of the atmospheres of
hot Jupiter exoplanets, much of the water these atmospheres contain
may be hidden beneath clouds. When the spectral modulation due to
H$_{2}$O is measured in units of the atmospheric scale height, we find
that the average H$_{2}$O-induced modulation is 1.8 $\pm$ 0.5~$H_{s}$ and never
exceeds 2.9~$H_{s}$. The shape of the spectral modulation is also
inconsistent with extremely low H$_{2}$O abundances and suggests that
a cloud layer may obscure a significant portion of the otherwise
observable atmosphere. We also find
that the sample of H$_{2}$O-hosting planets possesses a representative
spectral shape dominated by opacity due to H$_{2}$O. The fact that the
individual spectra coherently average to a consistent shape for the
H$_{2}$O-hosting planets, which is reproducible with simple forward
models, gives confidence that there is a representative spectrum for
at least a significant portion of the hot Jupiter exoplanet
population.


\section{Acknowledgements}
We are grateful to Dr. Yan Betremieux, Dr. David Crisp, Dr. Wladimir Lyra, and Dr. Yuk L. Yung for several insightful discussions on our analysis and suggestions for edits on the paper draft. 

This research is based on observations made with the NASA/ESA Hubble Space Telescope, obtained from the Data Archive at the Space Telescope Science Institute, which is operated by the Association of Universities for Research in Astronomy, Inc., under NASA contract NAS 5-26555.

We thank the JPL Exoplanet Science Initiative for partial support of this work. The research was carried out at the Jet Propulsion Laboratory, California Institute of Technology, under a contract with the National Aeronautics and Space Administration. $\copyright$ 2015. All rights reserved.

This research has made use of the Exoplanet Orbit Database
and the Exoplanet Data Explorer at exoplanets.org.

We thank the anonymous referee for their helpful comments.

\bibliography{references}

\begin{thebibliography}{36}
\expandafter\ifx\csname natexlab\endcsname\relax\def\natexlab#1{#1}\fi

\bibitem[{{Barman}(2008)}]{barman08}
{Barman}, T.~S. 2008, \apjl, 676, L61

\bibitem[{{Benneke}(2015)}]{benneke15}
{Benneke}, B. 2015, ArXiv e-prints

\bibitem[{{Brown}(2001)}]{brown01}
{Brown}, T.~M. 2001, \apj, 553, 1006

\bibitem[{{Deming} {et~al.}(2013){Deming}, {Wilkins}, {McCullough}, {Burrows},
  {Fortney}, {Agol}, {Dobbs-Dixon}, {Madhusudhan}, {Crouzet}, {Desert},
  {Gilliland}, {Haynes}, {Knutson}, {Line}, {Magic}, {Mandell}, {Ranjan},
  {Charbonneau}, {Clampin}, {Seager}, \& {Showman}}]{deming13}
{Deming}, D., {Wilkins}, A., {McCullough}, P., {Burrows}, A., {Fortney}, J.~J.,
  {Agol}, E., {Dobbs-Dixon}, I., {Madhusudhan}, N., {Crouzet}, N., {Desert},
  J.-M., {Gilliland}, R.~L., {Haynes}, K., {Knutson}, H.~A., {Line}, M.,
  {Magic}, Z., {Mandell}, A.~M., {Ranjan}, S., {Charbonneau}, D., {Clampin},
  M., {Seager}, S., \& {Showman}, A.~P. 2013, \apj, 774, 95

\bibitem[{{Ehrenreich} {et~al.}(2014){Ehrenreich}, {Bonfils}, {Lovis},
  {Delfosse}, {Forveille}, {Mayor}, {Neves}, {Santos}, {Udry}, \&
  {S{\'e}gransan}}]{ehrenreich14}
{Ehrenreich}, D., {Bonfils}, X., {Lovis}, C., {Delfosse}, X., {Forveille}, T.,
  {Mayor}, M., {Neves}, V., {Santos}, N.~C., {Udry}, S., \& {S{\'e}gransan}, D.
  2014, \aap, 570, A89

\bibitem[{{Fraine} {et~al.}(2014){Fraine}, {Deming}, {Benneke}, {Knutson},
  {Jord{\'a}n}, {Espinoza}, {Madhusudhan}, {Wilkins}, \& {Todorov}}]{fraine14}
{Fraine}, J., {Deming}, D., {Benneke}, B., {Knutson}, H., {Jord{\'a}n}, A.,
  {Espinoza}, N., {Madhusudhan}, N., {Wilkins}, A., \& {Todorov}, K. 2014,
  \nat, 513, 526

\bibitem[{{Grillmair} {et~al.}(2008){Grillmair}, {Burrows}, {Charbonneau},
  {Armus}, {Stauffer}, {Meadows}, {van Cleve}, {von Braun}, \&
  {Levine}}]{grillmair08}
{Grillmair}, C.~J., {Burrows}, A., {Charbonneau}, D., {Armus}, L., {Stauffer},
  J., {Meadows}, V., {van Cleve}, J., {von Braun}, K., \& {Levine}, D. 2008,
  \nat, 456, 767

\bibitem[{{Han} {et~al.}(2014){Han}, {Wang}, {Wright}, {Feng}, {Zhao},
  {Fakhouri}, {Brown}, \& {Hancock}}]{han14}
{Han}, E., {Wang}, S.~X., {Wright}, J.~T., {Feng}, Y.~K., {Zhao}, M.,
  {Fakhouri}, O., {Brown}, J.~I., \& {Hancock}, C. 2014, \pasp, 126, 827

\bibitem[{{Huitson} {et~al.}(2013){Huitson}, {Sing}, {Pont}, {Fortney},
  {Burrows}, {Wilson}, {Ballester}, {Nikolov}, {Gibson}, {Deming}, {Aigrain},
  {Evans}, {Henry}, {Lecavelier des Etangs}, {Showman}, {Vidal-Madjar}, \&
  {Zahnle}}]{huitson13}
{Huitson}, C.~M., {Sing}, D.~K., {Pont}, F., {Fortney}, J.~J., {Burrows},
  A.~S., {Wilson}, P.~A., {Ballester}, G.~E., {Nikolov}, N., {Gibson}, N.~P.,
  {Deming}, D., {Aigrain}, S., {Evans}, T.~M., {Henry}, G.~W., {Lecavelier des
  Etangs}, A., {Showman}, A.~P., {Vidal-Madjar}, A., \& {Zahnle}, K. 2013,
  \mnras, 434, 3252

\bibitem[{{Knutson} {et~al.}(2014a){Knutson}, {Benneke}, {Deming}, \&
  {Homeier}}]{knutson14a}
{Knutson}, H.~A., {Benneke}, B., {Deming}, D., \& {Homeier}, D. 2014a, \nat,
  505, 66

\bibitem[{{Knutson} {et~al.}(2014b){Knutson}, {Dragomir}, {Kreidberg},
  {Kempton}, {McCullough}, {Fortney}, {Bean}, {Gillon}, {Homeier}, \&
  {Howard}}]{knutson14b}
{Knutson}, H.~A., {Dragomir}, D., {Kreidberg}, L., {Kempton}, E.~M.-R.,
  {McCullough}, P.~R., {Fortney}, J.~J., {Bean}, J.~L., {Gillon}, M.,
  {Homeier}, D., \& {Howard}, A.~W. 2014b, \apj, 794, 155

\bibitem[{{Kreidberg} {et~al.}(2014a){Kreidberg}, {Bean}, {D{\'e}sert},
  {Benneke}, {Deming}, {Stevenson}, {Seager}, {Berta-Thompson}, {Seifahrt}, \&
  {Homeier}}]{kreidberg14a}
{Kreidberg}, L., {Bean}, J.~L., {D{\'e}sert}, J.-M., {Benneke}, B., {Deming},
  D., {Stevenson}, K.~B., {Seager}, S., {Berta-Thompson}, Z., {Seifahrt}, A.,
  \& {Homeier}, D. 2014a, \nat, 505, 69

\bibitem[{{Kreidberg} {et~al.}(2014b){Kreidberg}, {Bean}, {D{\'e}sert}, {Line},
  {Fortney}, {Madhusudhan}, {Stevenson}, {Showman}, {Charbonneau},
  {McCullough}, {Seager}, {Burrows}, {Henry}, {Williamson}, {Kataria}, \&
  {Homeier}}]{kreidberg14b}
{Kreidberg}, L., {Bean}, J.~L., {D{\'e}sert}, J.-M., {Line}, M.~R., {Fortney},
  J.~J., {Madhusudhan}, N., {Stevenson}, K.~B., {Showman}, A.~P.,
  {Charbonneau}, D., {McCullough}, P.~R., {Seager}, S., {Burrows}, A., {Henry},
  G.~W., {Williamson}, M., {Kataria}, T., \& {Homeier}, D. 2014b, \apjl, 793,
  L27

\bibitem[{{Kreidberg} {et~al.}(2015){Kreidberg}, {Line}, {Bean}, {Stevenson},
  {Desert}, {Madhusudhan}, {Fortney}, {Barstow}, {Henry}, {Williamson}, \&
  {Showman}}]{kreidberg15}
{Kreidberg}, L., {Line}, M.~R., {Bean}, J.~L., {Stevenson}, K.~B., {Desert},
  J.-M., {Madhusudhan}, N., {Fortney}, J.~J., {Barstow}, J.~K., {Henry}, G.~W.,
  {Williamson}, M., \& {Showman}, A.~P. 2015, ArXiv e-prints

\bibitem[{{Lecavelier Des Etangs} {et~al.}(2008){Lecavelier Des Etangs},
  {Pont}, {Vidal-Madjar}, \& {Sing}}]{lecavelierdesetangs08}
{Lecavelier Des Etangs}, A., {Pont}, F., {Vidal-Madjar}, A., \& {Sing}, D.
  2008, \aap, 481, L83

\bibitem[{{Line} {et~al.}(2013b){Line}, {Knutson}, {Deming}, {Wilkins}, \&
  {Desert}}]{line13b}
{Line}, M.~R., {Knutson}, H., {Deming}, D., {Wilkins}, A., \& {Desert}, J.-M.
  2013b, \apj, 778, 183

\bibitem[{{Line} {et~al.}(2014){Line}, {Knutson}, {Wolf}, \& {Yung}}]{line14}
{Line}, M.~R., {Knutson}, H., {Wolf}, A.~S., \& {Yung}, Y.~L. 2014, \apj, 783,
  70

\bibitem[{{Line} {et~al.}(2013a){Line}, {Wolf}, {Zhang}, {Knutson}, {Kammer},
  {Ellison}, {Deroo}, {Crisp}, \& {Yung}}]{line13a}
{Line}, M.~R., {Wolf}, A.~S., {Zhang}, X., {Knutson}, H., {Kammer}, J.~A.,
  {Ellison}, E., {Deroo}, P., {Crisp}, D., \& {Yung}, Y.~L. 2013a, \apj, 775,
  137

\bibitem[{{Madhusudhan} {et~al.}(2014){Madhusudhan}, {Crouzet}, {McCullough},
  {Deming}, \& {Hedges}}]{madhusudhan14}
{Madhusudhan}, N., {Crouzet}, N., {McCullough}, P.~R., {Deming}, D., \&
  {Hedges}, C. 2014, \apjl, 791, L9

\bibitem[{{Mandell} {et~al.}(2013){Mandell}, {Haynes}, {Sinukoff},
  {Madhusudhan}, {Burrows}, \& {Deming}}]{mandell13}
{Mandell}, A.~M., {Haynes}, K., {Sinukoff}, E., {Madhusudhan}, N., {Burrows},
  A., \& {Deming}, D. 2013, \apj, 779, 128

\bibitem[{{Markwardt}(2009)}]{markwardt09}
{Markwardt}, C.~B. 2009, in Astronomical Society of the Pacific Conference
  Series, Vol. 411, Astronomical Data Analysis Software and Systems XVIII, ed.
  D.~A. {Bohlender}, D.~{Durand}, \& P.~{Dowler}, 251

\bibitem[{{McCullough} {et~al.}(2014){McCullough}, {Crouzet}, {Deming}, \&
  {Madhusudhan}}]{mccullough14}
{McCullough}, P.~R., {Crouzet}, N., {Deming}, D., \& {Madhusudhan}, N. 2014,
  \apj, 791, 55

\bibitem[{{Mendez}(2014)}]{mendez14}
{Mendez}, A. 2014, in Habitable Worlds Across Time and Space, proceedings of a
  conference held April 28-May 1 2014 at the Space Telescope Science Institute.
  Online at href=''http://www.stsci.edu/institute/conference/habitable-worlds",
  id.32, 32

\bibitem[{{Morley} {et~al.}(2012){Morley}, {Fortney}, {Marley}, {Visscher},
  {Saumon}, \& {Leggett}}]{morley12}
{Morley}, C.~V., {Fortney}, J.~J., {Marley}, M.~S., {Visscher}, C., {Saumon},
  D., \& {Leggett}, S.~K. 2012, \apj, 756, 172

\bibitem[{{Pont} {et~al.}(2008){Pont}, {Knutson}, {Gilliland}, {Moutou}, \&
  {Charbonneau}}]{pont08}
{Pont}, F., {Knutson}, H., {Gilliland}, R.~L., {Moutou}, C., \& {Charbonneau},
  D. 2008, \mnras, 385, 109

\bibitem[{{Ranjan} {et~al.}(2014){Ranjan}, {Charbonneau}, {D{\'e}sert},
  {Madhusudhan}, {Deming}, {Wilkins}, \& {Mandell}}]{ranjan14}
{Ranjan}, S., {Charbonneau}, D., {D{\'e}sert}, J.-M., {Madhusudhan}, N.,
  {Deming}, D., {Wilkins}, A., \& {Mandell}, A.~M. 2014, \apj, 785, 148

\bibitem[{{Schwartz} \& {Cowan}(2015)}]{schwartz15}
{Schwartz}, J.~C., \& {Cowan}, N.~B. 2015, \mnras, 449, 4192

\bibitem[{{Seager} \& {Sasselov}(2000)}]{seager00}
{Seager}, S., \& {Sasselov}, D.~D. 2000, \apj, 537, 916

\bibitem[{{Sing} {et~al.}(2009){Sing}, {D{\'e}sert}, {Lecavelier Des Etangs},
  {Ballester}, {Vidal-Madjar}, {Parmentier}, {Hebrard}, \& {Henry}}]{sing09}
{Sing}, D.~K., {D{\'e}sert}, J.-M., {Lecavelier Des Etangs}, A., {Ballester},
  G.~E., {Vidal-Madjar}, A., {Parmentier}, V., {Hebrard}, G., \& {Henry}, G.~W.
  2009, \aap, 505, 891

\bibitem[{Sing {et~al.}(2016)Sing, Fortney, Nikolov, Wakeford, Kataria, Evans,
  Aigrain, Ballester, Burrows, Deming, D{\'e}sert, Gibson, Henry, Huitson,
  Knutson, Etangs, Pont, Showman, Vidal-Madjar, Williamson, \& Wilson}]{sing16}
Sing, D.~K., Fortney, J.~J., Nikolov, N., Wakeford, H.~R., Kataria, T., Evans,
  T.~M., Aigrain, S., Ballester, G.~E., Burrows, A.~S., Deming, D., D{\'e}sert,
  J.-M., Gibson, N.~P., Henry, G.~W., Huitson, C.~M., Knutson, H.~A., Etangs,
  A. L.~d., Pont, F., Showman, A.~P., Vidal-Madjar, A., Williamson, M.~H., \&
  Wilson, P.~A. 2016, Nature, 529, 59

\bibitem[{{Sing} {et~al.}(2015){Sing}, {Wakeford}, {Showman}, {Nikolov},
  {Fortney}, {Burrows}, {Ballester}, {Deming}, {Aigrain}, {D{\'e}sert},
  {Gibson}, {Henry}, {Knutson}, {Lecavelier des Etangs}, {Pont},
  {Vidal-Madjar}, {Williamson}, \& {Wilson}}]{sing15}
{Sing}, D.~K., {Wakeford}, H.~R., {Showman}, A.~P., {Nikolov}, N., {Fortney},
  J.~J., {Burrows}, A.~S., {Ballester}, G.~E., {Deming}, D., {Aigrain}, S.,
  {D{\'e}sert}, J.-M., {Gibson}, N.~P., {Henry}, G.~W., {Knutson}, H.,
  {Lecavelier des Etangs}, A., {Pont}, F., {Vidal-Madjar}, A., {Williamson},
  M.~W., \& {Wilson}, P.~A. 2015, \mnras, 446, 2428

\bibitem[{{Swain} {et~al.}(2014){Swain}, {Line}, \& {Deroo}}]{swain14}
{Swain}, M.~R., {Line}, M.~R., \& {Deroo}, P. 2014, \apj, 784, 133

\bibitem[{{Swain} {et~al.}(2008){Swain}, {Vasisht}, \& {Tinetti}}]{swain08}
{Swain}, M.~R., {Vasisht}, G., \& {Tinetti}, G. 2008, \nat, 452, 329

\bibitem[{{Tinetti} {et~al.}(2007){Tinetti}, {Vidal-Madjar}, {Liang},
  {Beaulieu}, {Yung}, {Carey}, {Barber}, {Tennyson}, {Ribas}, {Allard},
  {Ballester}, {Sing}, \& {Selsis}}]{tinetti07}
{Tinetti}, G., {Vidal-Madjar}, A., {Liang}, M.-C., {Beaulieu}, J.-P., {Yung},
  Y., {Carey}, S., {Barber}, R.~J., {Tennyson}, J., {Ribas}, I., {Allard}, N.,
  {Ballester}, G.~E., {Sing}, D.~K., \& {Selsis}, F. 2007, \nat, 448, 169

\bibitem[{{Wakeford} {et~al.}(2013){Wakeford}, {Sing}, {Deming}, {Gibson},
  {Fortney}, {Burrows}, {Ballester}, {Nikolov}, {Aigrain}, {Henry}, {Knutson},
  {Lecavelier des Etangs}, {Pont}, {Showman}, {Vidal-Madjar}, \&
  {Zahnle}}]{wakeford13}
{Wakeford}, H.~R., {Sing}, D.~K., {Deming}, D., {Gibson}, N.~P., {Fortney},
  J.~J., {Burrows}, A.~S., {Ballester}, G., {Nikolov}, N., {Aigrain}, S.,
  {Henry}, G., {Knutson}, H., {Lecavelier des Etangs}, A., {Pont}, F.,
  {Showman}, A.~P., {Vidal-Madjar}, A., \& {Zahnle}, K. 2013, \mnras, 435, 3481

\bibitem[{{Wilkins} {et~al.}(2014){Wilkins}, {Deming}, {Madhusudhan},
  {Burrows}, {Knutson}, {McCullough}, \& {Ranjan}}]{wilkins14}
{Wilkins}, A.~N., {Deming}, D., {Madhusudhan}, N., {Burrows}, A., {Knutson},
  H., {McCullough}, P., \& {Ranjan}, S. 2014, \apj, 783, 113

\end{thebibliography}

\end{document}